\documentclass[useAMS,usenatbib]{mn2e}
\usepackage{graphicx}
\usepackage{amssymb}
\usepackage{amsmath}
\usepackage{natbib}

\newcommand{\re}{\operatorname{Re}}

\newcommand{\be}{\begin{equation}}
\newcommand{\ee}{\end{equation}}
\newcommand{\bea}{\begin{eqnarray}}
\newcommand{\eea}{\end{eqnarray}}

\newcommand{\bz}{\boldsymbol{z}}
\newcommand{\bu}{\boldsymbol{u}}
\newcommand{\bv}{\boldsymbol{v}}
\newcommand{\bn}{\boldsymbol{n}}
\newcommand{\alm}{{a_{\ell m}}}

\title[Multipole Vectors Probability Density for Gaussian CMB]
{Probability Density of the Multipole Vectors for a Gaussian Cosmic Microwave Background}
\author[Mark R Dennis, Kate Land]{Mark R Dennis$^{1}$\thanks{E-mail: mark.dennis@physics.org}, 
Kate Land$^{2}$\thanks{E-mail: krl@astro.ox.ac.uk}\\
$^{1}$School of Mathematics, University of Southampton, Highfield, Southampton SO17 1BJ, UK\\
$^{2}$Oxford Astrophysics, Physics, DWB, Keble Road, Oxford, OX1 3RH, UK}

\begin{document}

\date{Accepted xxx. Received xxx; in original form xxx}

\pagerange{\pageref{firstpage}--\pageref{lastpage}} \pubyear{2007}

\maketitle

\label{firstpage}

\begin{abstract}
We review Maxwell's multipole vectors, and elucidate some of their mathematical properties, with emphasis on the application of this tool to the cosmic microwave background (CMB). 
In particular, for a completely random function on the sphere (corresponding to the statistically isotropic Gaussian model of the CMB), we derive the full probability density function of the multipole vectors.
This function is used to analyze the internal configurations of the third-year Wilkinson Microwave Anisotropy Probe quadrupole and octopole, and we show the observations are consistent with the Gaussian prediction. 
A particular aspect is the planarity of the octopole, which we find not to be anomalous.
\end{abstract}


\section{Introduction}

The cosmic microwave background (CMB) is a unique cosmological probe. 
It is a window onto the early Universe, a backlight on the more recent, and it defines our observable Universe. 
It is therefore our best tool for investigating predictions of inflation theories, and cosmological scale properties of space, such as statistical isotropy. 
Predictions from inflation, such as homogeneity, isotropy, and the existence of Gaussian fluctuations, currently form the basis for our fundamental framework, with consequences for all other cosmological studies. It is therefore essential that we test these predictions, with appropriate tools.
Current and future observations of the cosmic microwave background (CMB) provide us with a wealth of cosmological information, but they also confront us with many challenges. 
The large data-sets require novel data analysis and map-making techniques if we are to thoroughly exploit the data of its scientific information.

Since its launch in 2001, NASA's Wilkinson Microwave Anisotropy Probe (WMAP) has observed the CMB over the full-sky, in five frequency bands between 22 and 90 GHz. Multiple frequency coverage has enabled reliable separation of the Galactic foreground signal from the CMB anisotropy, 
producing spectacular results~\citep{Spergel:2003cb,Spergel:2006hy}. 
The observations provide further support for the $\Lambda$CDM concordance cosmological model, and are remarkably consistent with a range of independent observations such as the clustering of large-scale structure, and luminosity distances from 
Supernovae observations. However, some interesting features have been reported, on large-scales, which are anomalous with respect to our standard cosmological framework of a Gaussian statistically isotropic CMB. If an alternative framework is required, then this has far reaching consequences for the interpretation of all cosmological observations, and for the determination of cosmological parameter constraints~\citep{asymmparams}.

On multipole scales $\ell\lesssim 60$ there is an asymmetrical distribution of power on the CMB sky, with more power in the southern ecliptic hemisphere~\citep{erik07,erik1,hbg,dondon}. 
Similar asymmetrical signals have been seen with the bispectrum, $n$-point correlation functions, spherical wavelets, Minkowski functionals, genus statistic, and local curvature~\citep{Land:2004bs, Npoint, asymmwave, Mink, genus, asymmcurv}.

A separate `anomaly' involves just the largest scales, with $\ell=2,3$ correlated with each other and with the dipole direction~\citep{Copi:2003kt, Copi:2005ff,virgo,otzh:significance,Copi:2007}. 
Furthermore, $\ell=4,5$ are \emph{also} seen to align with this `Axis of Evil' direction~\citep{AOE}. 
Independently, $\ell = 3$ are reported to be particularly planar~\citep{TOH,otzh:significance,Schwarz:2004gk}, and the power on these scales is marginally low~\citep{lowCl}.  
The multipole vectors, described below, are an important tool in these studies.

However, the interpretation of these results is unclear. 
Different methods are employed, and various statistical measures invented, without systematically examining underlying mathematical properties of the representation. 
It is also not clear if some `anomalies' are connected and thus over-reported, and a more systematic basis for tests of non-Gaussianity of the CMB is required. 
The usual harmonic $n$-point correlation functions, the  bispectrum, trispectrum, etc., are unattractive as they are computationally intensive, abstract with relation to the underlying properties of the map, mutually dependent, contain redundant information, and are only meaningful under the assumption of statistical isotropy.

In this paper we revisit the multipole vectors, and study their mutual probability distribution for Gaussian maps.
As an example of the general results, we consider the planarity of the octopole, but not the possibility of correlations in direction between different multipoles.
The multipole vectors are a geometric representation of the spherical CMB map which is explicitly rotationally invariant, and whose probability distributions can be calculated for various Gaussian and non-Gaussian distributions of the CMB.
They were introduced into cosmology and numerically analyzed by \cite{Copi:2003kt}, and were subsequently studied by \cite{lm:wmap, lm:multipole, dennis:majorana, dennis:gaussiansphere, lrey, katz, weeksMV}, among others. 
More recent astrophysical applications of the multipole vectors include further analysis of the CMB data~\citep{Rakic}, characterizing the peculiar velocity field of supernovae~\citep{SNMV}, and quantifying systematic effects of dipole straylight contamination~\citep{DSC}.

Most of the multipole vector CMB analysis has involved a comparison of various derived quantities with the equivalent values from numerical Monte Carlo simulations; here, we present an alternative approach, based on analytic evaluation of the multipole vector probability distributions for statistically isotropic Gaussian random maps. 
We compute the full probability density function for the total multipole vector configuration, and use this to predict analytically the statistics of the multipole vector configuration.
In Section~\ref{MVintro} we introduce the multipole vectors and describe their numerical evaluation using polynomial root-finding algorithms (through the Majorana polynomial formalism); in Section~\ref{prob} we review their properties for Gaussian random maps, and present the analytic Gaussian multipole vector probability density (calculated in the Appendix). 
In Section~\ref{results}, these general statistical results are compared to the recent WMAP data at the largest scales, namely the quadrupole and octopole.
In particular, we will describe an analytically tractable measure of planarity for the octopole, and demonstrate the WMAP data is very close to the Gaussian prediction. 
We conclude with a discussion in Section~\ref{discuss}.

\section{The Multipole Vectors}\label{MVintro}

It is natural to represent data on the direction sphere (such as the cosmic microwave background) through the Fourier expansion
\be
   f(\theta,\phi) = \sum_{\ell \ge 0} \sum_{m = -\ell}^{\ell} \alm Y_{\ell m}(\theta,\phi),
   \label{eq:ylmdecomp}
\ee
for the complex spherical harmonic functions
\be
   Y_{\ell m}(\theta,\phi) \equiv \sqrt{\frac{2\ell+1}{4\pi}\frac{(l-m)!}{(\ell+m)!}} P_{\ell}^m(\cos \theta)e^{im\phi},
\ee
where $m = -\ell, -\ell + 1, ..., -1, 0, 1, ..., \ell - 1, \ell$, and $P_{\ell}^m$ is an associated Legendre polynomial. 
$Y_{\ell m}(\theta,\phi) = Y_{\ell m}(\bn)$ is the angular portion of the general solution to Laplace's equation in spherical polar coordinates, and $\ell$ corresponds to the total angular momentum quantum number, and corresponds to an angular size of $\theta \sim 180^\circ/\ell$.
All information in the map is now contained in the spherical harmonic coefficients $a_{\ell m}$, with reality of $f$ enforced by the condition 
\be 
   a_{\ell, -m}=(-1)^m a^*_{\ell m}
   \label{eq:selfinverse}.
\ee 
The expansion (\ref{eq:ylmdecomp}) for a fixed $\ell$ (`multipole') has $2\ell + 1$ real degrees of freedom.

Following the Copernican principle -- that there should be no special observers in the Universe -- statistical isotropy (SI) is a natural assumption to make about the Universe. 
Assuming SI for all observers further implies homogeneity, and thus it is common for us to adopt the most general homogeneous and isotropic framework in cosmology, namely the Friedmann-Robertson-Walker metric.
However, alternative frameworks exist, such as the anisotropic Bianchi models or the inhomogeneous Lema{\^i}tre-Tolman-Bondi models. 
Preferred directions would also exist if the Universe had a non-trivial topology, or contained large-scale magnetic fields.
SI of the CMB map is equivalent to the property that no quantity depends on any particular frame of reference -- the map is statistically invariant to global rotations.
Deviation from SI in the CMB observations would imply exotic physics in the early Universe or, perhaps more realistically, residual contamination by anisotropic foregrounds (e.g. emissions by Galactic diffuse components) or by residual systematic effects.

Such statistical properties of the spherical map can easily be conferred to the $a_{\ell m}$s. 
We assume that any cosmic probability distribution of the $\alm$s is ergodic, that is, spatial averaging is equivalent to ensemble averaging, and such averaging is denoted $\left< \bullet \right>$.
Therefore, the 2-point correlation function can be written in terms of the $\alm$ coefficients, 
\be 
   \left<f({\bn}_1)f({\bn}_2)\right> = \sum_{\ell_{1,2}, m_{1,2}} 
   \left< a_{\ell_1 m_1} a_{\ell_2 m_2} \right> Y_{\ell_1 m_1}({\bn}_1)Y_{\ell_2 m_2}({\bn}_2),
\ee 
and the action of a rotation $R$ on the spherical harmonics involves the Wigner $\mathcal{D}$-function (rotation matrix element)
\be 
   R[Y_{\ell m}({\bn})]=\sum_{m'} \mathcal{D}^\ell_{m' m}Y_{\ell m'}.
\ee
The orthogonality condition of the $\mathcal{D}$-function, 
\be
   \sum_m (-1)^{m_2-m} \mathcal{D}^{\ell_1}_{m_1 m}\mathcal{D}^{\ell_1}_{-m_2 -m}=\delta_{m_1 m},
\ee
implies that $\left<f({\bf n}_1)f({\bf n}_2)\right>$ is rotationally invariant (i.e. statistically isotropic) if and only if 
\be 
   \left<a_{\ell_1 m_1}a^*_{\ell_2 m_2} \right> = \delta_{\ell_1 \ell_2}\delta_{m_1 m_2}C_{\ell_1}
   \label{eq:amgauss},
\ee 
where $C_\ell$ is the \emph{angular power spectrum}.

The issue of Gaussianity is separate from SI, and if the CMB fluctuations are Gaussian, then all the information in the map is contained in the 2-point correlation function, or equivalently in the angular power spectrum $C_\ell$ as defined in (\ref{eq:amgauss}); higher order functions -- which otherwise probe non-Gaussianity -- depend only on the 2-point correlation function, or are zero.
SI is a consequence of Gaussianity, but there are other, non-Gaussian, statistically isotropic ensembles; this will be discussed further in Section \ref{prob}.

\begin{figure}
\begin{centering}
\includegraphics[width=6.4cm]{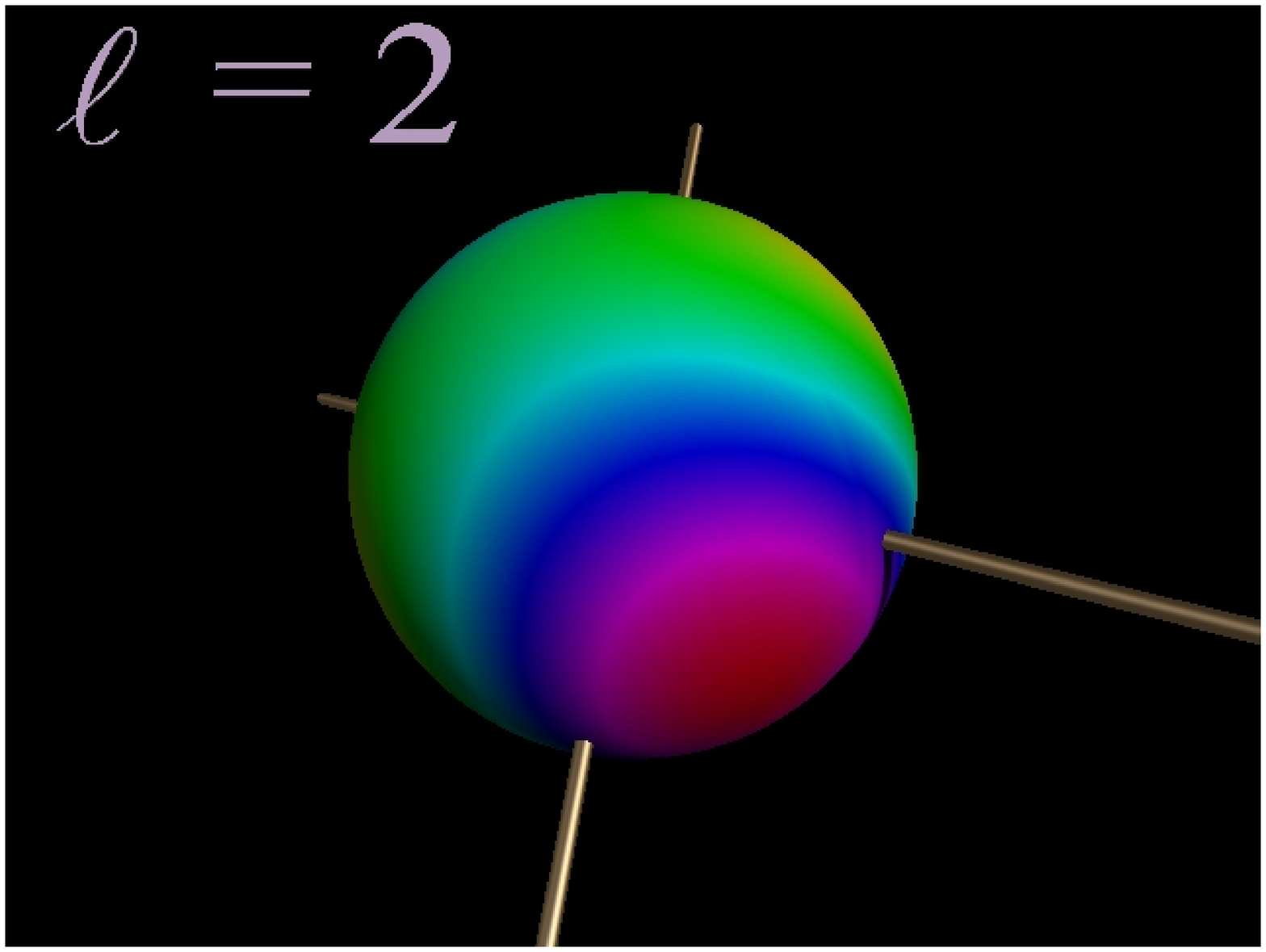}
\includegraphics[width=6.4cm]{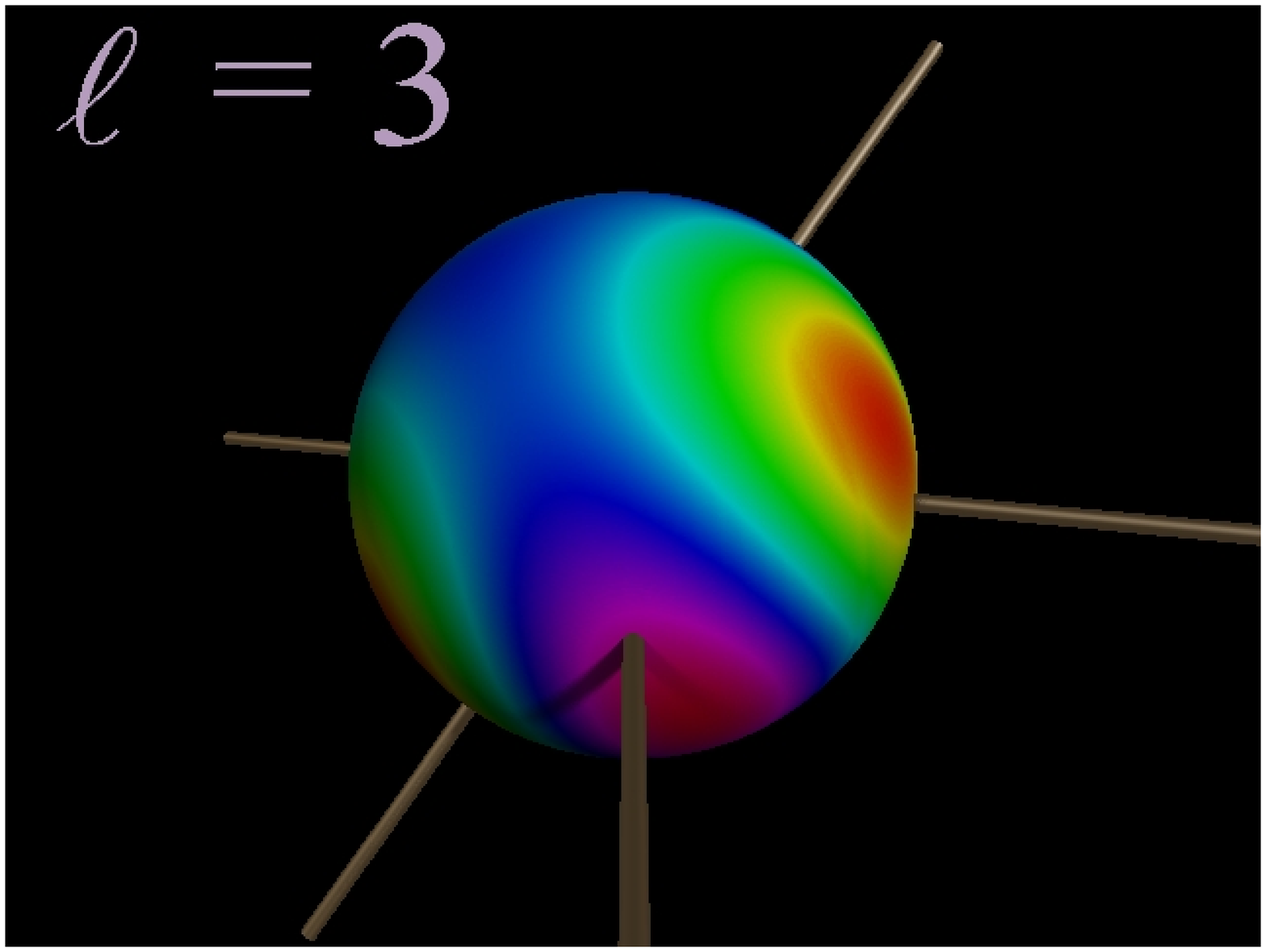}
\includegraphics[width=6.4cm]{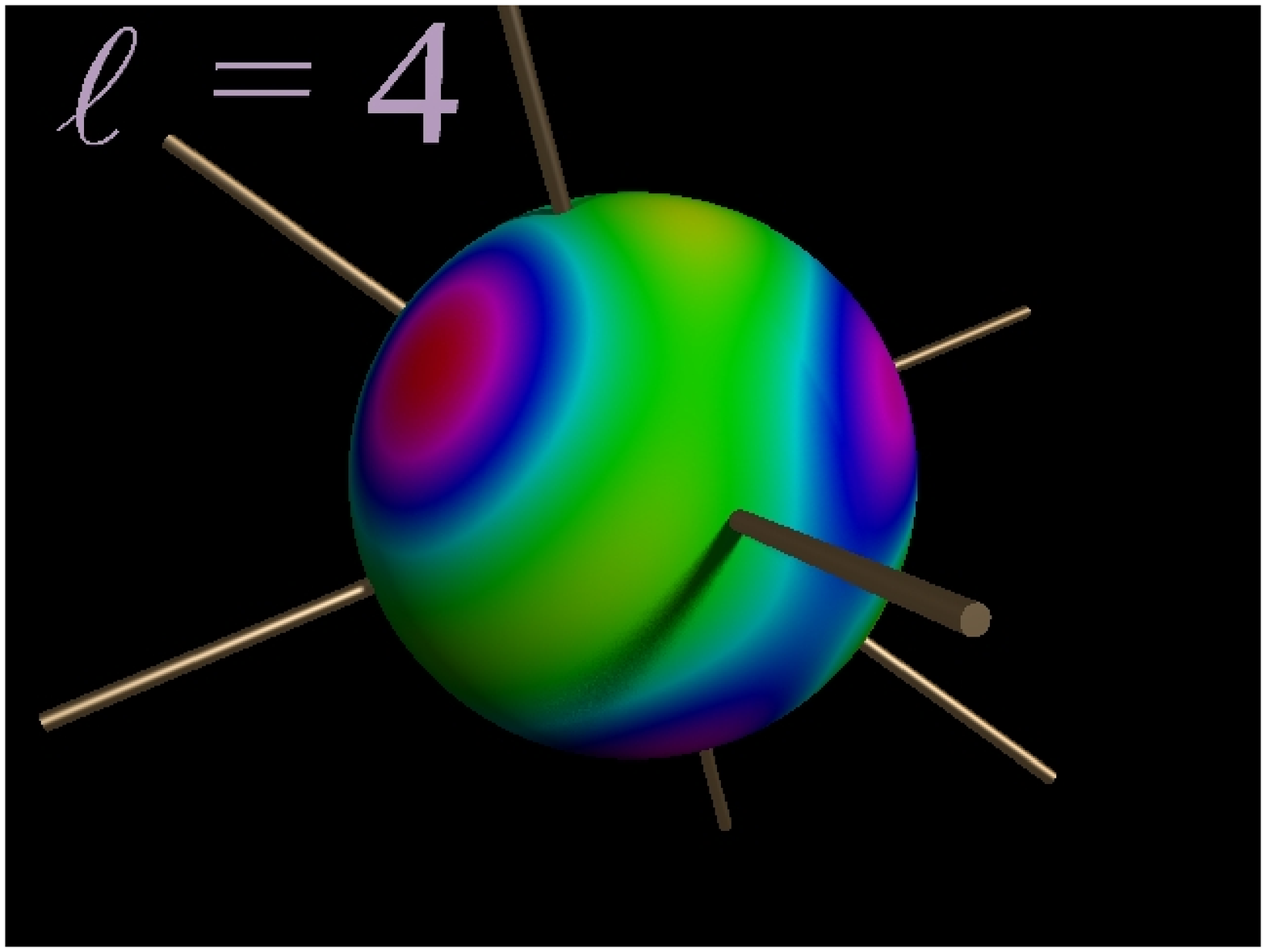}
\includegraphics[width=6.4cm]{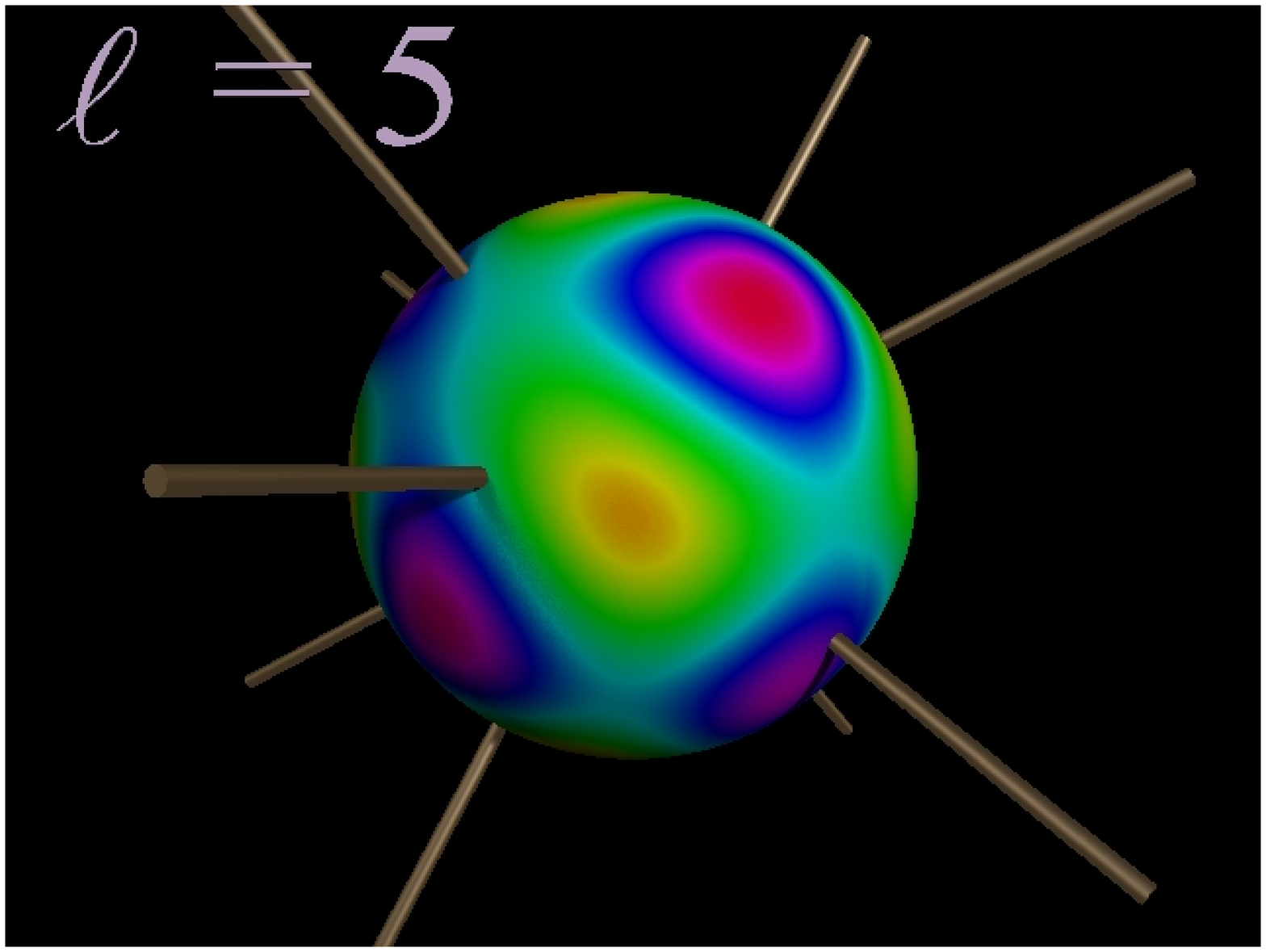}

\end{centering}
\caption{Representation of the multipole vectors for the WMAP3 multipoles $\ell=2,3,4,5$. 
The colour map on the spheres represents the underlying WMAP temperature data.}\label{MVfig}
\end{figure}

An alternative representation to (\ref{eq:ylmdecomp}) of a function on a sphere is via the \emph{multipole vectors} (MVs). 
For a real multipole of order $\ell$, represented
\be
   f_{\ell}(\theta,\phi) = \sum_{m = -\ell}^{\ell} \alm Y_{\ell m}(\theta,\phi),
   \label{felldecomp}
\ee
with coefficients satisfying (\ref{eq:selfinverse}) (i.e.~$f_{\ell}(x,y,z)$ is an eigenfunction of the Laplacian on the unit sphere ($x^2 + y^2 + z^2 = 1$) with eigenvalue $-\ell(\ell+1)$), there exist $\ell$ unit vectors ${\bv}_1,\ldots,{\bv}_\ell$ such that 
\be
   f_{\ell}(x,y,z)= A D_{{\bv}_1}\ldots D_{{\bv}_\ell}\frac{1}{r},
   \label{dder} 
\ee
where $D_{{\bv}_i}= {\bv}_i\cdot\nabla$ is a directional derivative operator, $r=({x^2+y^2+z^2})^{1/2}$ and $A$ is a real constant. 
A more useful form of the representation is given by~\cite{dennis:majorana}, 
\be
   f_{\ell}({\bn}) = B({\bn}\cdot{\bv}_1)\ldots({\bn}\cdot{\bv}_\ell) + r^2 T
   \label{easyvec}, 
\ee 
where $B$ is another real constant, and $T$ is a scalar, given by the contraction of a certain Cartesian tensor of rank $\ell - 2$ with unit vector $\bn$, which is fully determined by the $2\ell$ components of the ${\bv}_i$ and is completely traceless and symmetric~\citep{zz:maxwell}. 
A graphical representation of the $\ell = 2, 3, 4$ components of the WMAP temperature data of the CMB, together with their multipole vectors, is shown in Figure~\ref{MVfig}.

A simple way of connecting the two multipole representations (\ref{felldecomp}) and (\ref{easyvec}) is using stereographic projection of the direction sphere (as the Riemann sphere) into the complex plane; thus the point on the sphere with coordinates $\theta, \phi$ is stereographically mapped to the complex number $e^{i \phi} \tan(\theta/2)$.
The stereographic images of the multipole vectors appear as the roots of the Majorana polynomial (spin coherent state)~\citep{dennis:majorana} (also see~\cite{hst:cmb}),
\be
   p(\zeta) = \sum_{m=-\ell}^{\ell} \alm \mu_{\ell m} \zeta^{m+\ell},
   \label{eq:majorana}
\ee
where $\mu_{\ell m}$ is a numerical factor, 
\be
   \mu_{\ell m} \equiv (-1)^m \sqrt{\frac{(2\ell)!}{(\ell - m)! (\ell+m)!}}.
   \label{eq:mudef}
\ee
Thus $p(\zeta)$ is the complex polynomial acquired by replacing, in the expression (\ref{felldecomp}), $Y_{\ell m}$ with $\mu_{\ell m} \zeta^{\ell + m}$.
By the fundamental theorem of algebra, $p(\zeta)$ can be factorized:
\be
   p(\zeta) = a_{\ell} \prod_{i=1}^{2\ell} (\zeta - \zeta_i) = a_{\ell} \prod_{i=1}^{\ell} (\zeta - \zeta_i)(\zeta + 1/\zeta_i^{\ast}).
   \label{eq:majfactor}
\ee
The vectors corresponding to the roots $\zeta_i$ form antipodal pairs since $a_{\ell -m} = (-1)^m a_{\ell m}^{\ast}$, implying that if $\zeta_i$ is a root, so is its antipode on the Riemann sphere $-1/\zeta_i^{\ast}$~\citep{dennis:majorana}, and we label the roots such that $\zeta_{i+\ell} = -1/\zeta_i^{\ast}$, i.e. $\zeta_i, \zeta_{i+\ell}$ are antipodal.
The roots can be written in terms of modulus and argument,
\be
   \zeta_i = \tan(\theta_i/2) \exp(i \phi_i),
   \label{eq:roots}
\ee
with $\theta_i, \phi_i$ the spherical coordinates of the $i$th multipole vector.

The representations (\ref{eq:majorana}), (\ref{eq:majfactor}) provide a simple method for computing the MVs, since the problem is reduced to finding the complex roots of a polynomial of order $2\ell$. 
Alternative methods include those of~\cite{Copi:2003kt} which use an iterative algorithm to solve (\ref{easyvec}), and a similar approach was previously outlined by~\cite{me:poles}, albeit in a different physical context. 
Obviously, all methods have limited numerical accuracy, due to the global nature of the problem. 
We note that numerical root-finding algorithms run into fundamental difficulties for high-order polynomials, due to problems of ill-conditioning and the limited numerical accuracy of the polynomial coefficients (i.e.~the $\alm$ coefficients); with 2 decimal place accuracy in the coefficients, practical root-finding is probably limited to $\ell \lesssim 20$ \citep{acton:numerical}. 
It is harder to locate the source of numerical error in the iterative approach. 
On account of these numerical problems, MVs have been studied in the CMB for lower multipoles, particularly for $\ell \le 10$~\citep{Copi:2003kt,lm:wmap,weeksMV,Rakic,DSC,SNMV,Copi:2005ff}.


The MVs are, in fact, directors (`headless vectors'), since their sign is globally undefined (a $\pm$ transform on any pair of $\bv_i$ leaves (\ref{dder}), (\ref{easyvec}) unchanged). 
Furthermore, just like roots of a polynomial, they are not ordered amongst themselves. 
Counting the degrees of freedom we see that for a multipole $\ell$, there are $2\ell$ degrees of freedom in the $\ell$ unit vectors, and one in the constant $A$ in (\ref{dder}) (or $B$ in (\ref{easyvec}), agreeing with the $2\ell + 1$ freedoms in the $\alm$ coefficients).

A simple example of the multipole vectors is those of a pure, real spherical harmonic $\re Y_{\ell m}$. 
Of the $\ell$ MVs, $\ell - |m|$ are aligned with the $z$-axis (in the $Y_{\ell m}$ frame) and the remaining $|m|$ vectors lie in the $xy$-plane, with vertices forming a regular $2m$-polygon. 
Since this system of vectors rotates with the function, it is easy to identify pure harmonic modes in \emph{any} frame of reference; the distribution of the $\alm$s of a pure mode in an arbitrary frame of reference is certainly not so transparent. 
This also applies to arbitrary multipoles, whose configuration of MVs rotates directly with rotation of the function, unlike the coefficients $\alm$ (which change according to the appropriate $\mathcal{D}$-functions).
This property makes the MVs an extremely useful tool for studying the statistical isotropy of the CMB - their configuration and statistics do not depend on the frame of reference. 
This avoids the problems inherent in manipulating the $\alm$s directly, and of having to find complicated rotationally invariant functions (bispectrum, trispectrum, etc.) which are hard to connect to underlying properties of the map and increasingly computationally intensive to compute.
A drawback of the MVs is the problem of their numerical determination, described above.

The multipole vector formalism was apparently first introduced by Maxwell~\citep{maxwell}, and is discussed in several classic mathematical textbooks~\citep{hobson:spherical,ch:methods1} (also see the references in \cite{dennis:majorana}).
It has been applied in studies of quantum chemistry~\citep{me:poles} and continuum mechanics~\citep{backus:geometric,zz:maxwell}.
Its generalization to arbitrary spin states (where $\ell$ may be a half-integer and the coefficients $\alm$ do not necessarily satisfy (\ref{eq:selfinverse})) leads to the important canonical decomposition of totally symmetric spinors~\citep{pr:spinors1} and the stellar representation of spin coherent states and SU(2) polynomials~\citep{bacry:orbits,bbl:quantum,schupp:lieb}. 
In cosmology, the MV method has received considerable attention, from mathematicians exploring features of the formalism~\citep{dennis:gaussiansphere,katz,lrey,weeksMV,hst:cmb}, and from astrophysicists applying it to the WMAP data of the CMB~\citep{Copi:2005ff,Copi:2003kt,lm:wmap}.

In the following section, we will discuss the implications of different statistical assumptions of the $\alm$s on the probability distribution of the multipole vectors.


\section{Gaussian model}\label{prob}

A candidate ensemble for the $\alm$ coefficients is `Gaussian', that is, they are independent, identically distributed complex Gaussian random variables. 
Gaussianity of the CMB fluctuations is a generic prediction of simple inflation theories, because during slow-roll inflation the near flatness of the potential results in each Fourier mode of the inflaton perturbation evolving independently. 
This property translates to the density perturbations during reheating, and the central limit theorem implies their sum (i.e.~into real space) will be Gaussian. 
Furthermore, the probability distribution for each Fourier mode is already expected to be Gaussian (if the perturbation is in the vacuum state).
These models also predict that the Fourier modes have a near scale-invariant power spectrum. 
Even in these simple models, deviations from Gaussianity are expected on small scales where the central limit theorem breaks down. 
More complicated inflation models~\citep{wands,nonG}, or alternative theories of structure formation such as topological defects~\citep{Battye}, can predict higher levels of non-Gaussianity.

Requiring SI alone is a far weaker assumption, that the probability density function (PDF) for the set of $\alm$s depends on rotationally invariant quantities (the angular power spectrum $C_\ell$, bispectrum, trispectrum, etc.), and does not fix the ensemble uniquely.
As described in the previous section, any probability density function (PDF) for the set of $\alm$ coefficients satisfying (\ref{eq:amgauss}) is statistically isotropic.
For instance, it is possible to construct statistically isotropic ensembles which have a fixed MV configuration that is randomly oriented (for instance, for $\ell = 2$, the two MVs have the same -- random -- direction, equivalent to $Y_{20}$ in some random frame).
A Gaussian PDF, in which the $\alm$s are independent and Gaussian (with identical variance for same $\ell$) clearly satisfies (\ref{eq:amgauss}), and is unique in the property that its 2-point correlation function (or equivalently, its power spectrum) determines the ensemble uniquely; statistically isotropic non-Gaussian ensembles, whilst satisfying (\ref{eq:amgauss}), have different $n$-point correlation functions (equivalently, averages of more than two $\alm$ coefficients are non-Gaussian).

Nevertheless, there are other statistically isotropic PDFs which have the same MV distribution as the Gaussian model -- using the MVs alone, it is impossible to distinguish Gaussianity uniquely from certain other ensembles. 
This wider class of ensembles is characterized by the property that the only quantity appearing in the PDF of the $\alm$s is $\hat{C}_\ell \equiv \sum_m |\alm|^2/(2\ell+1)$; equivalently, the $(2\ell+1)$-dimensional vector $(a_{\ell,-\ell}, ..., a_{\ell\ell})$ is invariant to all $(2\ell+1)$-dimensional unitary transformations preserving (\ref{eq:selfinverse}).
In \cite{dennis:gaussiansphere}, this wider class of ensembles was called {\em completely random}, as the statistics of the multipole $f_{\ell}$ depend only on the distribution of its normalization $\hat{C}_{\ell}$, and the MVs have the same statistics for any completely random ensemble, independent of the precise PDF of $\hat{C}_{\ell}$.
This may be understood from the polynomial representation (\ref{eq:majorana}): the MVs are distributed by the polynomial roots, whose distribution is independent of the any overall multiplicative factor of $p(\zeta)$, equivalent to $\hat{C}_{\ell}.$ 
Clearly, Gaussian distribution of the $\alm$s satisfies the condition to be completely random, and the PDF of $\hat{C}_{\ell}$ is that of the radius in a multidimensional isotropic Laplace distribution.
A non-Gaussian example of a completely random ensemble is that in which $\hat{C}_\ell$ is absolutely fixed (its PDF is a $\delta$-function); of course, such a distribution does not have any cosmological significance. 
Another completely random example is that of Gaussian and independent $\alm$s, apart from a cut forcing a maximum $\hat{C}_\ell$.
Such an ensemble was explored by~\cite{Rakic} -- indeed, they observe that in this case the MVs are indistinguishable from the independent Gaussian case. 
If, in addition to complete randomness, the $\alm$s are also assumed independent (i.e. the full PDF factorizes, $P(\{\alm\}) = P(a_{\ell,-\ell}) ... P(a_{\ell,\ell})$), then the completely random ensemble must be Gaussian (the argument is analogous to that of the Maxwell distribution in kinetic theory~\citep{dennis:gaussiansphere}).
Another feature of all completely random ensembles -- including the Gaussian -- is that the PDFs for separate multipoles are completely independent.
We will not examine this prediction against the CMB data in the present paper.

Perhaps counterintuitively, for a given $\ell,$ the MVs in the completely random case are \emph{not} statistically independent, but have a unique joint probability distribution, which we derive exactly below.
The derivation uses the fact that the MV distribution is equivalent to the probability distribution of the roots of the random Majorana polynomial (\ref{eq:majorana}), and is a special case of more general investigations of random polynomials~\cite[e.g.][]{hannay:exact,bbl:quantum,prosen:exact}, where independent Gaussian random coefficients are assumed, although not the condition (\ref{eq:selfinverse}).
In our analysis in Section \ref{results}, we will compare the MV distributions of the completely random model (including the Gaussian model) with that of independent MVs; although this latter does not have any obvious cosmological application, it is a non-Gaussian ensemble satisfying SI whose statistics are very simple to calculate analytically.

It has been previously shown~\citep{dennis:gaussiansphere,lm:multipole} that the 2-point function of the MVs exhibits quadratic repulsion (as usual for the roots of random polynomials and matrices~\citep{bbl:quantum}).
Given this repulsion, one may expect that the most probable MV distribution for fixed $\ell$ is at the vertices of the regular polyhedron with $2\ell$ vertices, if it exists.
If it does not exist, then there will be no unique maximal probability distribution (similar constructions for polynomials on the Riemann sphere are considered by~\cite{as:geometry}).

For any completely random distribution of the $\alm$ coefficients, including the Gaussian distribution, the probability distribution for the $\ell$ multipole vectors $\bv_1, ..., \bv_\ell$ can be found analytically, based on a transformation of the set of Gaussian-distributed coefficients $\{ \alm \}$ to the set of $\ell$ independent roots $\{ \zeta_i \}$ of the Majorana polynomial (\ref{eq:majorana}).
The full details of this calculation are given in the Appendix.
For the multipole of order $\ell$, the full joint probability density function of the $\ell$ multipole vectors $\{\bv_i\}$, assuming dependence only on $\hat{C}_{\ell}$, is given by
\bea
   P_{\ell}(\{\bv_i\}) &=& \frac{(2\ell - 1)!! \prod_{j=1}^{2\ell} j!}{ (2\pi)^{\ell} 2^{\ell^2} \ell! } \times   \label{eq:finalv}\\
   && \frac{\prod_{j,k=1, j<k}^{\ell} \sin^2\theta_{j k}}{\left(
   \sum_{\sigma \in S_{2\ell} } e^{i \alpha(\{\bv_i\},\sigma)/2} \prod_{j=1}^{2\ell} \cos\frac{\theta_{j,\sigma(j)}}{2}\right)^{\ell+1/2}},
   \nonumber
\eea
where $\theta_{jk}$ denotes the angle between the MVs labelled by $j$ and $k$, and in the denominator, summation is over permutations $\sigma$ in the set $S_{2\ell}$ of permutations on the $2\ell$ indices $j = 1, ..., 2\ell$; such a permutation gives rise to a system of piecewise-geodesic curves on the direction sphere (as each $\bv_j$ goes to $\bv_{\sigma(j)}$ to $\bv_{\sigma^2(j)},...$), and $\alpha(\{\bv_i\},\sigma)$ is the signed solid angle on the sphere enclosed by these curves. 
It should be noted that $\bv_{\ell + i} = - \bv_i$, and the term in the denominator involves the $2\ell$-element set $\{\bv_1, ..., \bv_{2\ell} \}$.

In practice, it is easier to work directly with the roots of the Majorana polynomial (stereographic images of the MVs in the complex plane), which have the probability distribution equivalent to (\ref{eq:finalv}),
\bea
   P_{\ell}(\{\zeta_i\}) &=& \frac{(2\ell-1)!! \prod_{j=1}^{2\ell} j!}{(2\pi)^{\ell} \ell!  \prod_{j=1}^{\ell} |\zeta_j|^2} \times\nonumber\\
   &&\quad\frac{\prod_{j,k=1, j<k}^{2\ell} |\zeta_j - \zeta_k|}{\left(\sum_{\sigma \in S_{2\ell} } \prod_{j=1}^{2\ell}  (1 + \zeta_j \zeta_{\sigma(j)}^{\ast})\right)^{\ell+1/2}},
   \label{eq:finalzeta}
\eea
The function (\ref{eq:finalv}), and especially its stereographic counterpart (\ref{eq:finalzeta}), bears some similarity to the full probability distributions of roots of other Gaussian random polynomials representing spin states~\citep{bbl:quantum,hannay:exact}: the discriminant of the polynomial in the numerator is divided by a combinatorial factor which mixes the roots nontrivially.

The functions $P_{\ell}(\{\bv_i\})$ and $P_{\ell}(\{\zeta_i\})$) contain all the statistical information of the directions of the MVs, and the functions for different $\ell$ are mutually independent; the function (\ref{eq:finalv}) is one of the main results of this paper.
The expressions (\ref{eq:finalv}) and (\ref{eq:finalzeta}) are unwieldy for high $\ell$, primarily because the sum in the denominator involves $(2\ell)!$ terms, most of which are nontrivial.

In \cite{dennis:gaussiansphere}, the 2-point multipole vector functions $\rho_{\ell}$ were calculated (corresponding to $P_{\ell}(\{\zeta_i\})$ only when $\ell = 2$). 
Normalized with respect to angle $\theta$ (between 0 and $\pi/2$), this is
\begin{eqnarray}
\lefteqn{   \rho_{\ell}(\theta)  =   \sin\theta \left(\ell(\ell-1) (1+\cos\theta)^2 D^{5/2}\right)^{-1} \times  }\nonumber \\
   && \left\{ (2\ell D - 4 b u v - (b^2 + v^2)(a - 1 - u^2) ) \right. \times \nonumber \\
   && (d D - 2 c u v (a + 1 - u^2) - (c^2+ a v^2)(a - 1 - u^2) ) \nonumber \\
   && + ( 2\ell D - 2 c u v - b u v (a + 1 -u^2) - v^2 (a - 1 + u^2) \nonumber \\
   && - b c (a - 1 - u^2))^2 + ( w D - 2 b c u - u v^2 (a + 1 - u^2) \nonumber \\
   && \left. - b v (a - 1 + u^2) - c v (a - 1 - u^2))^2 \right\}, \label{eq:rho21}
\eea
with $D = (a-1-u^2-2u)(a-1-u^2+2u)$ and
\bea
   &a = \sec^{4\ell}\frac{\theta}{2}, \; b = 2\ell \tan\frac{\theta}{2}, \; c = \ell \sin\theta \sec^{4\ell}\frac{\theta}{2},& \nonumber \\
   &d = 2\ell \sec^{4\ell-2}\frac{\theta}{2} \left(1+ 2\ell \tan\frac{\theta}{2}\right), \; u = \tan^{2\ell}\frac{\theta}{2},& \nonumber \\
   &v = -2\ell \tan^{2\ell-1}\frac{\theta}{2}, \; w = -2\ell(2\ell-1) \tan^{2\ell-2}\frac{\theta}{2}.&
   \label{eq:corrvals1}
\eea

For the remainder of the paper, we will compare the full joint probability $P_{\ell}$ and the 2-point $\rho_{\ell}$ for the lowest two nontrivial $\ell$-components of the CMB distribution, namely the quadrupole $\ell = 2$, and octopole $\ell = 3.$ 
This probability is shared by the Gaussian and other completely random ensembles; we will also compare the results for the ensemble of independent MVs. 
We will not be concerned here with issues of data analysis and error in determining the multipole vectors, but merely note that the full PDF enables the error in the multipole vector directions to be mapped directly to those in the $\alm$s.
The $\ell = 2$ case was considered before~\citep{lm:multipole,dennis:gaussiansphere}, but is included here for completeness; the $\ell = 3$ analysis is new.

\section{Comparison with CMB data, $\ell = 2,3$}\label{results}

For tests of statistical isotropy it is essential that we use full-sky maps, as a mask induces a relatively large anisotropic signal of its own. 
We therefore examine the `internal-linear-combination' (ILC) maps of the WMAP collaboration. 
The WMAP mission~\citep{Spergel:2003cb} produced full sky CMB maps from ten differencing assemblies (DAs).  
They also produced an ILC map, from combining smoothed frequency maps with weights chosen to minimize the rms fluctuations, using separate sets of weights for 12 disjoint sky regions. 
In the first-year data release the WMAP collaboration advised that the ILC map be used only as a visual tool. 
However, for the third-year release a thorough error analysis of the ILC map was performed, and a bias correction implemented~\citep{Hinshaw:2006ia}. 
The resulting third-year ILC map (herein WMAP3) is expected to be clean enough on scales $\ell\lesssim10$ to be used without a mask\footnote{WMAP data available from http://lambda.gsfc.nasa.gov}.

As a consistency check we compare the WMAP3 data with some third-party maps, in which different methods of removing the foreground signal were employed.
The third-year ILC map of~\cite{TOH} (TOH herein), calculated using an `internal' method (as WMAP), but with weights depending on scale in harmonic space as well as Galactic latitude. 
This is advantageous since different sources of contamination dominate at different scales - foregrounds at large scales, and noise at smaller scales. 
We also examine the map of~\cite{parkmap} (PARK herein), who employed an ILC method but fitting different weights over hundreds of disjoint regions rather than just 12. 
The final datset is that produced by~\cite{banday}, who reconstructed the low-multipoles from high-latitude WMAP data by means of the power equilization filter method. 
We use their V-band map (PE herein).

For $\ell = 2$, in the Galactic frame, the multipole vectors of the WMAP3 are 
\be
   \left( \begin{array}{r} -0.562 \\ 0.815 \\ 0.138 \end{array}\right),
   \left( \begin{array}{r} 0.971 \\ 0.048 \\ 0.234  \end{array}\right).
\ee
The angle between the quadrupole CMB vectors is $61.7^{\circ}$ (to 1 decimal place). 
The resulting vectors are very similar for our third-party maps, although the inter-angle varies somewhat: \{TOH, PARK, PE\} = $\{82.8^{\circ},74.5^{\circ},53.1^{\circ}\}$. 
Due to the low number of degrees of freedom, the low-$\ell$ multipoles are particularly sensitive to the various treatments of the galactic plane~\citep{lowl}. 
However, as we will see, this variation does not effect our conclusions.
Choosing the MVs (rather than their antipodes) such that the $z$-component of the MV is positive, the $\ell=3$ WMAP3 vectors are
\be
   \left( \begin{array}{r}0.905\\0.401\\0.145\end{array}\right),
   \left( \begin{array}{r}0.677\\-0.708\\0.202\end{array}\right),
   \left( \begin{array}{r} -0.051\\0.770\\0.636\end{array}\right).
   \label{eq:l3us}
\ee
We label these $\bu_0,\bu_1,\bu_2$ in sequence, and write down the three cosines between these vectors (which define the configuration uniquely)
\be
   \bu_0 \cdot \bu_1 =  0.358, \;
   \bu_1 \cdot \bu_2 = -0.451, \;
   \bu_2 \cdot \bu_0 =  0.353,
   \label{eq:l3cos}
\ee
corresponding to angles between the directors of $69.0^\circ, 63.2^\circ, 69.3^\circ$ respectively. 
The MV director angles for the third-party maps are
\bea
   \hbox{TOH} & : & 68.9^\circ, 66.9^\circ, 73.8^\circ \nonumber \\
   \hbox{PARK} & : & 65.4^\circ, 64.4^\circ, 73.5^\circ \nonumber \\
   \hbox{PE} & : & 64.5^\circ, 62.1^\circ, 65.4^\circ.
   \label{3rdparty}
\eea



We now analyze these MVs with the Gaussian predicted statistics of Section \ref{prob}.
In subsection \ref{2point}, we will discuss the 2-point MV correlation function and its comparison with the Gaussian 2-point function (\ref{eq:rho21}); in subsection \ref{3point}, we will discuss how the octopole data compares with $P_3(\{\bv_i\})$, and in subsection \ref{tau}, we will discuss a natural measure of non-planarity for the octopole whose distribution can be almost-analytically handled by the Gaussian MV distribution function $P_3$.

\subsection{2-point multipole vector correlation functions}\label{2point}

The 2-point correlation function probabilities were previously calculated in \cite{dennis:gaussiansphere}, and studied numerically in \cite{lm:multipole}.
When $\ell = 2$, this is
\be
   \rho_2(\cos\theta) = \frac{27(1-\cos^2\theta)}{(3+\cos^2\theta)^{5/2}}.
   \label{eq:rho2}
\ee
Since the angle $\theta$ between the two multipole axes is the only degree of freedom for the $\ell = 2$ configuration, this is also the joint PDF $P_2(\{\bv_1,\bv_2\})$ of the multipole vectors for the quadrupole, and can be explicitly shown from (\ref{eq:finalv}).
Normalizing over $0 \le \theta \le \pi/2$, therefore, we have 
\be
   P_2(\theta)=\frac{27\sin^3 \theta}{(3+\cos^2\theta)^{5/2}}.
\ee

\begin{figure}
\begin{center}
\includegraphics[bb=30 160 610 410,width=9.0cm]{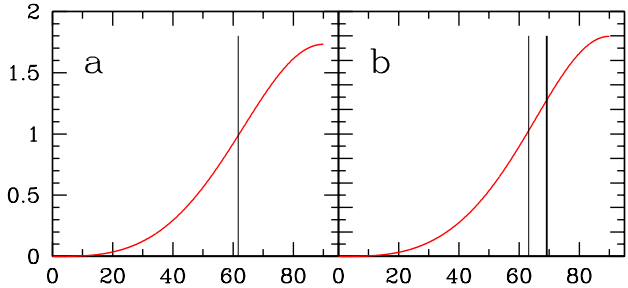}
\end{center}
    \caption{The 2-point correlation functions for Gaussian random multipoles, and the CMB WMAP3 data. 
    a) $\ell = 2$, with angular separations $\theta_{\mathrm{WMAP3}} = 61.7^{\circ};$ b) $\ell = 3$, with $\theta_{\mathrm{WMAP3}} = 63.2^{\circ}, 69.0^{\circ}, 69.3^{\circ}$.}
    \label{fig:rho}
\end{figure}

In Fig. \ref{fig:rho}a, $P_2$ is plotted together with the WMAP3 data $\theta_{\mathrm{WMAP3}}$ of $61.7^{\circ}$.
The most preferred angle is $90^{\circ}$, with a probability of $\mathrm{Prob}(\theta>80^\circ) = 0.29$, and a median of $\theta\approx 72.2^\circ$.  
For the WMAP3 data we find $\mathrm{Prob}(\theta<61.7^{\circ}) = 0.28$, giving no indication of any significant discord between the data and the Gaussian (completely random) prediction.
Similarly, the results from the third-party maps give are not significant as the lowest angle (for PE) finds $\mathrm{Prob}(\theta<53.1^{\circ}) = 0.16$.
Thus there appears to be nothing anomalous about the distribution of the multipole vectors for $\ell=2$.

When $\ell = 3$, the 2-point function is, from (\ref{eq:rho21}),
\be
\rho_3(X) = \frac{75 \sqrt{3} (1- X^2)(3+X^2)^2(88+67 X^2+34 X^4+3 X^6)}{((4-X+X^2)(4+X+X^2)(5+3X^2))^{5/2}}
\ee
with $X = \cos\theta$.
As with $P_2$, the maximum occurs when $\theta = 90^{\circ}$ -- when the vectors are orthogonal. 
The function is plotted in Fig. \ref{fig:rho}b, together with the data from (\ref{eq:l3cos}). 
We find, for the CMB data values, $\mathrm{Prob}(\theta<\theta_3)=0.29, 0.41, 0.42$ for $\theta_3=63.2,69.0,69.3^\circ$ respectively, with very similar results for the third-party maps.
Thus, again the 2-point MV correlation function does not highlight anything anomalous about the octopole, although we look to the full joint PDF considered in the next subsection for a more complete analysis.

\subsection{Full octopole Gaussian PDF}\label{3point}

Since the PDF of any statistically isotropic ensemble is invariant to the choice of reference frame for the MVs, it is convenient to rotate the system of MVs $\bv_0, \bv_1, \bv_2$ so that $\bv_0$ points in the $+z$-direction and $\bv_1$ lies in the $x,z$-plane

%

There are therefore three parameters which govern the internal geometry of the system: the $z$-coordinate $c_1$ of $\bv_1$ (i.e. $c_1 = \cos \theta_1$), the $z$-coordinate $c_2$ of $\bv_2$ (i.e. $c_2 = \cos \theta_2$) and the relative azimuthal angle $\phi$ between $\bv_1$ and $\bv_2$.
Our chosen standard coordinate system is therefore
\begin{equation}
   \bv_0 = \left( \begin{array}{c} 0 \\ 0 \\ 1 \end{array}\right), \:
   \bv_1 = \left( \begin{array}{c} s_1 \\ 0 \\ c_1  \end{array}\right), \:
   \bv_2 = \left( \begin{array}{c} s_2 \cos\phi \\ s_2 \sin\phi \\ c_2  \end{array}\right),
   \label{eq:l3vs}
\end{equation}
where $s_j = \sin \theta_j = \sqrt{1-c_j^2}, j = 1,2$. 
However, due to the antipodal nature of the MVs, it is not appropriate to define $c_1, c_2$ over the whole sphere, and according to the labelling convention of (\ref{eq:l3us}), $c_1, c_2 \ge 0$ (i.e. the MV axis is labelled by the vector in the northern hemisphere).

Using (\ref{eq:l3cos}), it is straightforward to determine that, for the CMB MV data (\ref{eq:l3us}) (with the same ordering),
\begin{eqnarray}
   c_{1,\mathrm{WMAP3}} & = & \bu_0 \cdot \bu_1 = 0.358,
   \nonumber \\
   c_{2,\mathrm{WMAP3}} & = & \bu_0 \cdot \bu_2 = 0.353,
   \label{eq:l3para}\\
   \phi_{\mathrm{WMAP3}} & = & \arg(\bu_0\times\bu_1)\cdot\left(\bu_0\times \bu_2 + i \bu_2 \right) 
   = - 2.293. 
   \nonumber
\end{eqnarray}
These are readily checked by verifying that the scalar products of (\ref{eq:l3vs}) agree with (\ref{eq:l3cos}), and that the scalar triple product is the same for the two sets of vectors.

The full $\ell = 3$ octopole PDF $P_3$ is given by (\ref{eq:finalv}) (actually, computed using (\ref{eq:finalzeta})), is significantly simplified by choosing the MVs in the standard coordinate system (\ref{eq:l3vs}), reducing to
\bea
   &&\hspace{-.5cm}P_3(c_1,c_2,\phi) = \label{eq:l3pdf}\\ &&\hspace{-.5cm}
   \frac{750\sqrt{3} s_1^2 s_2^2 (1 - (c_1 c_2 + s_1 s_2 \cos\phi)^2)}{\pi (5 + 3c_1^2 + 3c_2^2 + (c_1 c_2 + s_1 s_2 \cos\phi)(c_1 c_2 + 3 s_1 s_2 \cos\phi))^{7/2}}.
   \label{eq:p3}
\nonumber\eea
It is evident that $c_1$ and $c_2$ appear on an equal footing, as they should. 
Further, because the multipole vectors are not naturally ordered, this PDF does not depend on their ordering (although the exact equation for parameter $\phi$ in (\ref{eq:l3para}) depends on the $xy$-quadrant $\bu_2$ occupies in the rotated standard coordinate system).

$P_3$ is normalized, i.e.
\begin{equation}
   \int_0^1 d c_1 \int_0^1 d c_2 \int_0^{2\pi} d \phi \, P_3(c_1,c_2,\phi) = 1,
   \label{eq:p3norm}
\end{equation}
which is easily checked numerically.
Our choice of parameters is such that the probability density element $d c_1 \, d c_2 \, d \phi$ is uniform.
In the model in which the MVs are statistically isotropic and completely independent, the PDF for $c_1, c_2, \phi$ (over the same integration domain) would evidently be constant $1/2\pi$.

\begin{figure}
\begin{center}
\includegraphics[width=8cm]{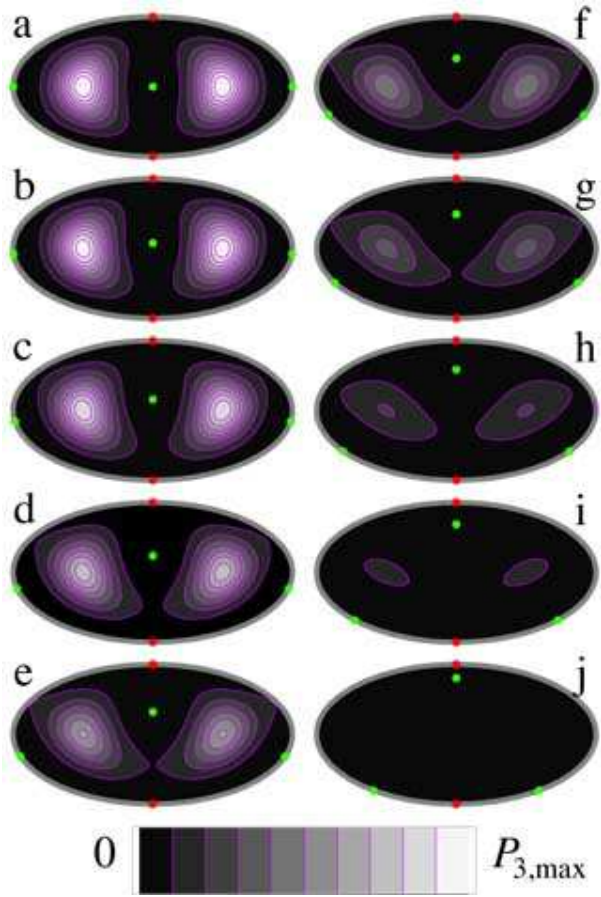}
\end{center}
    \caption{A representation of the 3-dimensional octopole vector PDF $P_3(c_1, c_2, \phi)$ from (\ref{eq:l3pdf}), in standard configuration.
    Each frame a-j represents a different position of $c_1 < 1$ (the green dot on the vertical axis), and is a plot over the $\theta_2, \phi$-sphere in Mollweide projection.
    Frame a has $c_1 = 0$, increasing in increments of $0.1$ to frame j, with $c_1 = 0.9$ (since $P_3(1,c_2, \phi) = 0$).
    The fixed MV axis in $\pm \hat{\bz}$ is represented by red dots at the north and south poles, and the opposite vector to $\bv_1$ is shown on each side of the plot (at $\phi = \pm\pi$) as green dots.
    For each frame (with fixed $c_1$), the shading represents the probability $P_3(c_1,c_2, \phi)$, that is, the position of the MV $\pm\bv_2$ (parameterized by $\theta_2, \phi$). 
    Contours are plotted at values of $P_{3,\mathrm{max}}/10$.    
    If the probability density function were uniform over $c_1, c_2, \phi$, the uniform density would be about the level of the lowest contour line.}
    \label{fig:l3fullpdf}
\end{figure}

The 3-dimensional distribution of $P_3(c_1,c_2, \phi)$ is shown in Fig. \ref{fig:l3fullpdf}. 
It is clear from the figure that the maximally probably distribution of the MV axes is mutually perpendicular, that is $c_1 = c_2 = 0$, $\phi = \pm\pi/2$, or $|\bv_0 \cdot (\bv_1 \times \bv_2)| = 1$.
This supports the previous statement that the unique maximal probability configuration exists when $2\ell$ is the number of vertices of a regular polyhedron, in this case an octahedron (whose six vertices are mutually perpendicular), and the MV configuration falls into the octahedral symmetry group (the octopole function with maximum probability ($\propto x y z \propto i Y_3^2 - i Y_3^{-2}$) has tetrahedral symmetry, not being inversion-symmetric).

This maximal probability occurs with probability density $6 \sqrt{3}/\sqrt{5} \pi \approx 1.48$, about $9$ times that for independent MVs.
The probability is low when the MVs are nearby, demonstrating repulsion as with the 2-point MV function.

\begin{figure}
\begin{center}
\includegraphics[width=8cm]{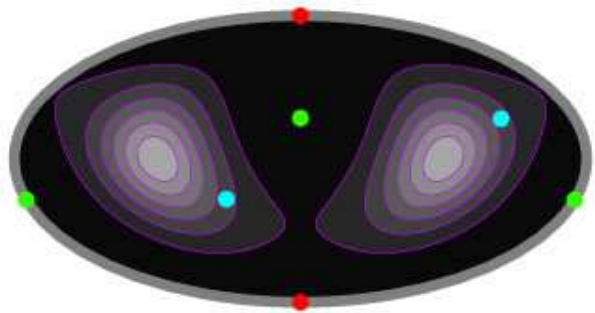}
\end{center}
    \caption{A representation, in the style of Fig. \ref{fig:l3fullpdf}, showing the CMB WMAP3 octopole, with the $\theta_2, \phi$ distribution with $c_1 = 0.358$ fixed. 
    The third WMAP3 multipole vector with coordinates $(\phi, c_2) = (48.6^{\circ},0.353)$ and its antipode are represented as a light blue pair of dots.
    The scale of the plot is the same as in Fig. \ref{fig:l3fullpdf}.}
    \label{fig:l3data}
\end{figure}

The MV directions of the octopole CMB data are represented in Fig. \ref{fig:l3data}, and this has probability density 0.362, which can be compared to the maximum (orthogonal configuration) with a probability density of 1.479, and the minimum (all aligned) with zero. 
We find that 87.4\% of the $\{c_1,c_2,\phi\}$ configuration space has a probability density less than 0.362, and thus the WMAP3 octopole MV configuration is not improbable.
The third-party maps (TOH, PARK, PE) have octopole configurations with probability densities $0.484, 0.380, 0.252$ respectively (with 91.5\%, 88.2\%, 81.2\% of configuration space with lower probability density).
Thus, according to the Gaussian multipole vector joint probability density function, there is nothing particularly anomalous about the $\ell = 3$ configuration of the multipole vectors, in agreement with~\cite{AOE, AOE2}.

This contrasts with the findings of~\cite{TOH,otzh:significance,Schwarz:2004gk}, in which the octopole data is interpreted to be remarkably planar. 
That is to say, the multipole vectors tend to lie equally spaced in a plane, or equivalently the octopole is dominated by $Y_{3,3}$ in some frame. 
These claims were re-examined in~\cite{Copi:2005ff}, where planarity was found with a reduced evidence of only $89\%$ Confidence Limit, using the same statistic.
Indeed the vectors do tend to lie in a plane, but this does not appear to be particularly anomalous. 
The reason for the disagreement originates in the particular statistical measure used, and various measures have been introduced in the literature, based upon their ease for numerical computation.  
Here, we develop an alternative, analytic approach based on the full joint probability distribution of the multipole vectors, which complements the other numerical approaches. 
As can be seen from Fig. \ref{fig:l3data}, the multipole vectors can have a significantly planar configuration while not being anomalous with respect to a Gaussian distribution. 
In the following subsection, we study octopole planarity further by calculating the statistics for a natural measure of MV planarity.
It should be noted that the discussions of the papers cited above, significance is added to the observed planarity of the octopole by its correlation with the orientation of the observed quadrupole; we make no comment on inter-multipole orientations here.

Comparing with the ensemble of the completely independent MVs model, whose probability density is constant $1/2\pi \approx 0.159$ (about the lowest contour line level in Figs. \ref{fig:l3fullpdf}, \ref{fig:l3data}) the CMB data is obviously more likely to originate from a Gaussian distribution of $\alm$ coefficients. 
The plots of Figs. \ref{fig:l3fullpdf}a and \ref{fig:l3data} are plotted spherically in Fig. \ref{fig:l3sphpdf}.

\begin{figure}
\begin{center}
\includegraphics[width=8cm]{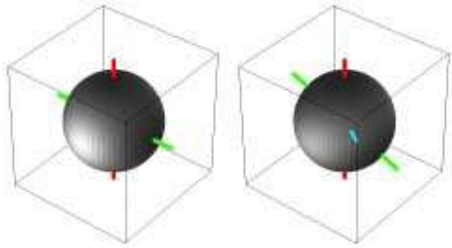}
\end{center}
    \caption{A further representation of Figs. \ref{fig:l3fullpdf}a and \ref{fig:l3data}, plotted now on the surface of the sphere.
    The multipole vector axes are represented by the same colours as before.}
    \label{fig:l3sphpdf}
\end{figure}


\subsection{Statistics of planarity of $\ell = 3$ multipole vectors}\label{tau}

Using the analytically-calculated probability density function for the three multipole vectors of the cosmic octopole, it is possible to predict exactly the non-planarity statistics of the Gaussian model.

As we described above, up to a global rotation, the octopole vectors have three degrees of freedom, and the full joint probability density function of these vectors is given for the Gaussian model in (\ref{eq:l3pdf}). 
We will compare this with the ensemble with completely independent MVs, whose joint PDF for $c_1,c_2,\phi$ is the constant $1/2\pi$.

An analytically natural measure of (non-)planarity of the multipole vectors $\bv_0, \bv_1, \bv_2$, is the magnitude of the scalar triple product
\begin{eqnarray}
   \tau & \equiv & | ( \bv_0 \times \bv_1) \cdot \bv_2 | \nonumber \\
   & = & | s_1 s_2 \sin \phi | 
   \label{eq:taudef}
\end{eqnarray} for $\bv_i$ given by (\ref{eq:l3vs}), where $s_i=\sqrt{1-c^2_i}$. 
The sign of the triple product has no meaning, as the ordering of the vectors is arbitrary, as is the choice of $\pm$ direction of the $\bv_j$. 
Geometrically, $\tau$ is the volume of the parallelepiped with edges $\bv_0, \bv_1, \bv_2$. 
Clearly it is $0$ if and only if the vectors are coplanar (linearly dependent), and it is 1 if and only if they are orthogonal.
$\tau$ does not directly determine the spacing of the MVs; in the data, not only are the MVs nearly planar, they are also quite uniformly spaced (their angles are all $\approx 60^{\circ}$).
Any planarity measure, including $\tau$, depends on the pairwise spacing of the MVs as well as their overall planarity.

In other studies, the normalized angular momentum dispersion, maximized over all directions, is used as the measure of octopole planarity \citep{otzh:significance,Copi:2005ff}. 
This quantity, which positively weights octopole configurations with equally spaced MVs ($a_{3,3}+a_{3,-3}$ in some frame), is rather hard to analyze using the analytical methods here; $\tau$ is used as an alternative planarity measure which is mathematically natural -- (\ref{eq:taudef}) is the simplest quantity to combine three vectors in three dimensions -- although equally spaced planar MVs, as opposed to any other planar configuration, are not particularly distinguished.
$\tau$ is properly a measure of non-planarity (being smaller the more planar the configuration), and we henceforth adopt this description.

In the Appendix, the PDF of $\tau$ is derived for the Gaussian model and the independent MVs model.
For the Gaussian model, the PDF is written down as a certain double integral which can be integrated numerically for each $\tau$.
In the independent MVs case, the PDF is simply
\begin{equation}
   P_{\mathrm{ind}}(\tau) = \arccos \tau.
   \label{eq:puncsig}
\end{equation}

\begin{figure}
\begin{center}
\includegraphics*[width=8cm]{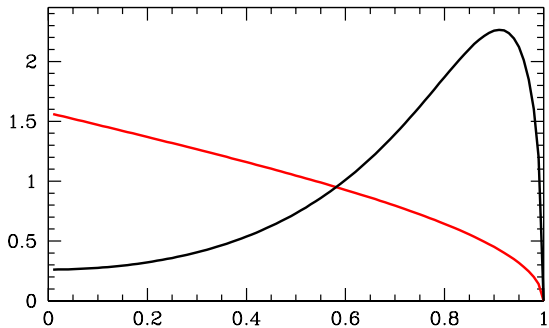}
\end{center}
    \caption{Plots of the probability density functions of the multipole vector non-planarity $\tau$ for the two ensembles: Gaussian model (black) and uncorrelated multipole vectors (red).}
    \label{fig:stpplot}
\end{figure}

Plots of the PDFs for $\tau$ for the two ensembles are presented in Fig. \ref{fig:stpplot}: the uncorrelated ensemble tends to have a lower (more planar) value of $\tau$ than the Gaussian model, on account of the repulsion of the MVs for the Gaussian ensemble.
The first two moments in each case are
\begin{eqnarray}
&&   \langle \tau \rangle_{\mathrm{ind}}  =   \pi/8 \approx 0.393,  \nonumber \\
&&   \langle \tau^2 \rangle_{\mathrm{ind}}  =  2/9 \approx 0.222,  \nonumber \\
&&   \langle \tau \rangle_{\mathrm{Gauss}}  =   0.686,  \nonumber \\
&&   \langle \tau^2 \rangle_{\mathrm{Gauss}}  =   0.529.  
   \label{eq:sigstats}
\end{eqnarray}
Thus, for the Gaussian model, although the configuration with maximum probability has orthogonal multipole vectors, corresponding to $\tau = 1$, the mean $\tau$ is less than 0.7.
The data for WMAP3 map, with octopole multipole vectors given by (\ref{eq:l3us}), has a non-planarity measure of
\begin{equation}
   \tau_{\mathrm{WMAP}} = 0.655,
   \label{eq:sigdat}
\end{equation}
which is surprisingly high given the literature on the apparent anomalous planarity of the octopole; for the Gaussian model this value is in very good agreement with the mean.
As described above, this difference arises in the choice of $\tau,$ rather than other measures, as planarity measure.

For the third-party maps, $\tau$ has values $0.749$ (TOH), $0.676$ (PARK) and $0.505$ (PE).

The full probability density (\ref{eq:l3pdf}) may also be used to investigate the behaviour of the multipole vectors if they are constrained to be planar; in the present parameterization, this is equivalent to the restriction $\phi = 0$ (the possibility that $\phi = \pi$ has been replaced by an extended range for $\theta_2,$ $-90^{\circ} \le \theta_2 \le 90^{\circ}$).
With these restrictions, the probability distribution (for integration with respect to $d c_1 \, d c_2$) is proportional to
\bea\lefteqn{
   P_{\mathrm{Gauss}}(\theta_1,\theta_2|\hbox{planar})    = }\label{eq:planarcond} \\ &\mbox{}& \hspace{1.0cm} A \frac{\sin^2\theta_1 \sin^2\theta_2 \sin^2(\theta_1-\theta_2)}{(9 + \cos 2\theta_1 + \cos 2\theta_2 + \cos2(\theta_1-\theta_2))^{7/2}}, \nonumber
\eea
where $A$ is a normalization constant.
This distribution is plotted in Fig. \ref{fig:planarprob}.
Although it is obvious that the most likely configuration with enforced planarity is with multipole vectors maximally distant (i.e. $\theta_1 = 60^{\circ}, \theta_2 = -60^{\circ}$), this distribution quantifies this precisely for the Gaussian model.

\begin{figure}
\begin{center}
\includegraphics*[width=8cm]{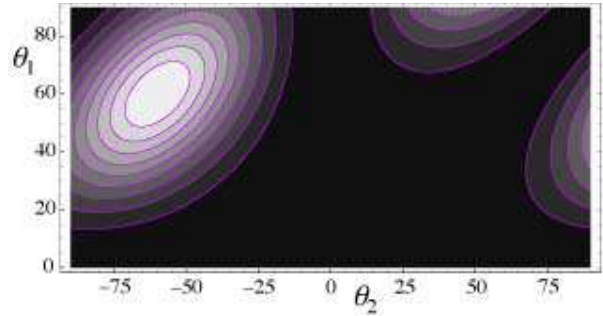}
\end{center}
    \caption{Probability distribution of the multipole vector configuration conditional on a planar arrangement.
    The vertical axis is $\theta_1$, the horizontal $\theta_2,$ both in degrees (with $0 \le \theta_1 \le 90^{\circ}$, $-90^{\circ} \le \theta_2 \le 90^{\circ}$ for all possible planar configurations).}
    \label{fig:planarprob}
\end{figure}

\section{Discussion}\label{discuss}

We have reviewed the multipole vector method of interpreting CMB maps, including an analytic computation of the full probability density function of the multipole vectors for the statistically isotropic Gaussian random model. 
We analyzed this function for $\ell = 2, 3$, and a comparison with data revealed that the third-year WMAP data is consistent with the Gaussian model, despite the apparent anomalous planarity of the octopole. 

A natural extension of this work is to provide a full PDF analysis of the higher multipoles; however, this is mathematically complicated. 
$P_3$ depends on three parameters, and therefore can be plotted (as in Fig. \ref{fig:l3fullpdf}) as a 1-parameter family of 2-dimensional maps.
The next multipole, $\ell = 4$, requires five parameters to specify the multipole vector configuration, and $P_5$ requires 7. 
In addition to this increase of parameters, the term in the denominator of the PDF (\ref{eq:finalv}), (\ref{eq:finalzeta}), summing over $(2\ell)!$ permutations of the MVs, becomes significantly more difficult to compute as $\ell$ increases. 
For $\ell>20$ the problem is exacerbated by the limited numerical accuracy of the various MV-finding algorithms. 
It is clear that as we extend the work to higher multipoles, statistics based on derived quantities, rather than the full PDF, will be necessary. 
Furthermore, we have not addressed the important issue of inter-$\ell$ correlations~\citep{AOE,AOE2}, or correlations between the CMB and external reference frames --  
as pointed out by~\cite{TOH,otzh:significance,Schwarz:2004gk,Copi:2005ff,Copi:2007}, the apparent plane of the octopole is strongly aligned with the orientation of the dipole. 
Such correlations are indicative of a breakdown of SI, independent from any assumption about the distribution of the $\alm$s. 

Thus, we defer to a future work a study where measures of anisotropy will be constructed from the MVs, in place of the full probability density function.
These anisotropy measures must be defined to be as orthogonal as possible so to avoid a repetition of information.
Also, the MVs will be used to systematically define a Cartesian frame for each multipole, so to probe the orientation and inter-$\ell$ correlations of the multipoles, and thus probe the issue of SI independently of any assumptions about Gaussianity.

It is fortuitous that the standard model for the CMB, namely Gaussian distributed $\alm$ coefficients, corresponds precisely to the case studied for spin coherent SU(2) polynomials~\citep{bbl:quantum,hannay:exact}, with the additional constraint (\ref{eq:selfinverse}).
This means that quantities such as the 2-point MV correlation function $\rho_2$, the full MV joint PDF $P_{\ell}$, and, in fact, any $N$-point MV correlation function~\citep{dennis:gaussiansphere} can be calculated exactly analytically using the methods of random polynomials.
This suggests the application to the CMB of other ideas from other fields this technique is used, such as random spins~\citep{hannay:exact}, Renyi-Werhl entropy~\citep{hst:cmb}, Skyrmion analysis~\citep{as:geometry}, and quantum chaos~\citep{bbl:quantum}, which may provide other powerful tools to analyze cosmological maps. 

One of the most difficult aspects of CMB analysis is, of course, that we are provided with only one sample function to test our statistics: the observed Universe. 
This fact is all the more apparent at the low multipoles where there are only a small number of degrees of freedom. 
As it is precisely these few data points that probe the largest scales in the Universe, it is essential that we hone our statistics to exploit the low-multipoles of the CMB as carefully as possible.


\section*{Acknowledgements}
We are grateful to John Hannay, Joao Magueijo, Glenn Starkman and James Vickers for discussions.
KRL is funded by the Glasstone Research Fellowship and Christ Church college, Oxford, MRD by a Royal Society University Research Fellowship.


\section*{Appendix}

\subsection*{Derivation of the full MV PDF (\ref{eq:finalv})}

The coefficients of any polynomial can be written in terms of the roots, the so-called {\em symmetric polynomials}.
In the present case, 
\begin{equation}
   p(\zeta) = a \exp\left(-i \sum_{i=1}^{\ell} \phi_i\right) \sum_{m = -\ell}^{\ell} (-1)^{\ell - m} s_{\ell - m} \zeta^{m + \ell},
   \label{eq:psym}
\end{equation}
where $s_n$ is the $n$th symmetric polynomial for the polynomial of order $2\ell$ (with $2\ell + 1$ coefficients, $n = 0, ..., 2\ell + 1$),
\begin{equation}
   s_n = \sum_{1 \le i_1 < i_2 < ... < i_n \le 2\ell} \prod_{k = 1}^n \zeta_{i_k},
   \label{eq:sidef}
\end{equation}
and the factor $a \exp\left(-i \sum_{i=1}^{\ell} \phi_i\right)$ is the coefficient for $\zeta^{2\ell}$.
Therefore
\begin{equation}
   \mu_m \alm = a \exp\left(-i \sum_{i=1}^{\ell} \phi_i\right) (-1)^{\ell - m} s_{\ell - m}.
   \label{eq:as}
\end{equation}
$a$ is a real number since the overall phase factor of $a_{\ell}$ is fixed.
Since $s_0 = 1,$ and $s_{2\ell} = \prod_{i=1}^{2\ell} \zeta_i = \exp\left(2i \sum_{i=1}^{\ell} \phi_i\right)$, then
\begin{equation}
   a_{-\ell} 
   = (-1)^{\ell} a \exp\left(-i \sum_{i=1}^{\ell} \phi_i\right) s_{2\ell} 
   = (-1)^{\ell} a_{\ell}^{\ast}.
   \label{eq:alphase}
\end{equation}

By hypothesis, the coefficients $a_0, a_1, ..., a_{\ell}$ are independent and identically distributed Gaussians ($a_0$ is real, the others complex).
The correctly normalized total probability distribution is therefore  
\begin{eqnarray}
   P(\{\alm\}) & = & \frac{1}{\sqrt{2} \pi^{\ell+1/2}} \exp\left(-\frac{a_0^2}{2} - \sum_{m=1}^{\ell} |\alm|^2\right) 
   \nonumber \\
   & = & \frac{1}{\sqrt{2} \pi^{\ell+1/2}} \exp\left(-\frac{1}{2}\sum_{m=-\ell}^{\ell} \frac{a^2 |s_{m+\ell}|^2}{\mu_m^2}\right)
   \nonumber \\
   & = & \frac{1}{\sqrt{2} \pi^{\ell+1/2}} \exp\left(-\frac{1}{2}\frac{a^2}{2(2\ell)!} \Sigma \right).
   \label{eq:fullapdf}
\end{eqnarray}
In the second line, $|\alm|^2$ has been replaced with $(|\alm|^2 + |a_{-m}^2|^2)/2 = a^2 (|s_{\ell+m}|^2 +|s_{\ell-m}^2|^2)/2\mu_m^2$ using (\ref{eq:as}). 
In the final line, the quantity $\Sigma$ has been introduced, where
\begin{equation}
   \Sigma \equiv \sum_{i = 0}^{2\ell} i! (2\ell-i)! |s_i|^2 = \sum_{\sigma \in S_{2\ell}} \prod_{i=1}^{2\ell} (1 + \zeta_i \zeta_{\sigma(i)}^{\ast}),
   \label{eq:sigmadef}
\end{equation}
where the second equality, in which $\Sigma$ is realized as a sum over permutations $\sigma$ over the $2\ell$ roots, was observed by~\cite{hannay:exact}.
The quantity $\hat{C}_{\ell} \equiv \sum_{m=-\ell}^{\ell} |\alm|^2/(2\ell+1),$ described in Section \ref{prob}, equals $a^2 \Sigma/2(2\ell)!$ by (\ref{eq:fullapdf}).
The PDF of any completely random ensemble depends only on this quantity.

In order to write the probability density (\ref{eq:fullapdf}) as a distribution over the multipole vector root configuration, it is necessary to find the appropriate Jacobian $\{\alm\} \to \{\zeta_i\}$.
Using standard methods (e.g.~\cite{bbl:quantum}) it is straightforward to show that
\begin{equation}
   d^{2\ell+1}\{\alm\} = \frac{a^{2\ell}}{\prod_{m=-\ell}^{\ell} |\mu_m |}\frac{\mathcal{D}}{\prod_{j=1}^{\ell} |\zeta_i|^2} d a \, d^{2\ell}\{\zeta_j\}.
   \label{eq:trans}
\end{equation}
where $\mathcal{D}$ is the square root modulus of the polynomial discriminant,
\begin{equation}
   \mathcal{D} = \prod_{i,j=1, i<j}^{2\ell} |\zeta_i - \zeta_j|.
   \label{eq:disc}
\end{equation}

The final joint PDF for the multipole vectors is the integral of this quantity with respect to $a$ ($-\infty < a < \infty$) (using the Gaussian probability density (\ref{eq:fullapdf}).
It must be divided through by $\ell!$ (since the $\ell$ independent roots are indistinguishable), and $2^{\ell}$ (since $\zeta_j, \zeta_{j+\ell}$ are indistinguishable for $j = 1, ..., \ell$):
\begin{eqnarray}
   P_{\ell}(\{\zeta_i\}) 
   & = & \frac{1}{\ell! (2\pi)^{\ell+1/2}} \frac{\mathcal{D}}{\prod_{m = -\ell}^{\ell} |\mu_m| \prod_{j=1}^{\ell} |\zeta_j|^2} \times \nonumber \\
   & &  \int d a \, a^{2\ell} \, \exp\left(-\frac{a^2}{2(2\ell)!} \Sigma\right)
   \label{eq:final1} \\
   & = & \frac{(\ell-1/2)!}{\ell! \pi^{\ell+1/2}} \frac{(2\ell)!^{\ell+1/2}\mathcal{D}}{\Sigma^{\ell+1/2} \prod_{m = -\ell}^{\ell} |\mu_m| \prod_{j=1}^{\ell} |\zeta_j|^2}.
   \nonumber
\end{eqnarray}
Rearranging the numerical prefactor, $(\ell - 1/2)!/\pi^{\ell+1/2} = (2\ell - 1)!/(2\pi)^{\ell}$, and $(2\ell)!^{\ell+1/2} /\prod_{m = -\ell}^{\ell} |\mu_m| = \prod_{j = 1}^{2\ell} j!$.
Finally, substituting $\Sigma$ via (\ref{eq:sigmadef}) gives (\ref{eq:finalzeta}).

In order to transform from the stereographic plane to the direction sphere, it is necessary to multiply by the stereographic Jacobian factor $2^{-\ell} \prod_{j = 1}^{\ell}(1 + |\zeta_j |^2)^2$ (this factor differs from the usual stereographic Jacobian factor by $2^{\ell}$, since the labelling MVs are in the northern hemisphere, meaning that $|\zeta_j|\le 1$ for $j = 1, ..., \ell$).
Substituting the roots by (\ref{eq:roots}), with appropriate interpretation on the direction sphere~\citep{hannay:exact} gives the joint probability density function (\ref{eq:finalv}).

\subsection*{Derivation of the $\tau$ PDF}

The probability density function for $\tau$ is found by substitution of $\phi \to \tau$ by
\begin{equation}
   \phi = \arcsin\left(\frac{\tau}{s_1 s_2}\right).
   \label{eq:phiid}
\end{equation}
The Jacobian of this transformation is
\begin{equation}
   \frac{d \phi}{d \tau} = \frac{1}{\sqrt{s_1^2 s_2^2 - \tau^2}}
   \label{eq:sigjac}
\end{equation}
The derivation of the PDF for $\tau$ in the independent MVs model is easy, as the PDF is constant with respect to the volume element $d c_1 \, d c_2 \, d\phi$.
The PDF can therefore be calculated from the $c_1, c_2, \phi$ normalization integral, with $\phi$ substituted by $\tau$ using (\ref{eq:phiid}) and (\ref{eq:sigjac}), then integrating $c_1$ and $c_2.$ 
This integral is
\begin{equation}
   1 = \frac{2}{\pi}\int_0^1 d c_1 \int_0^1 d c \int_0^{\sqrt{(1-c_1^2)(1-c_2^2)}} d \tau \frac{1}{\sqrt{(1-c^2)(1-c_1^2) - \tau^2}}.
   \label{eq:uncorpdf1}
\end{equation}
Without changing the integrand, the order of integration is changed to $c_2$, then $c_1$, then $\tau$ last, with $0 \le c_2 \le \sqrt{1 - \tau^2/(1-c_1^2)}$, $0 \le c_1 \le 1 - \tau^2$, then $0 \le \tau \le 1$.
Finally, $c_1$ and $c_2$ are rescaled in order that the volume of integration is again a cube:
\begin{equation}
   1 = \frac{2}{\pi}\int_0^1 d \tau \int_0^1 d c_1 \int_0^1 d c_2 \frac{\sqrt{1-\tau^2}}{\sqrt{(1-c_1^2)}\sqrt{(1-c_2^2(1-\tau^2))}}.
   \label{eq:uncorpdf3}
\end{equation}
$c_1$ and $c_2$ may now be integrated directly,
\begin{equation}
   1 = \int_0^1 d \tau \arccos \tau,
   \label{eq:uncorpdf4}
\end{equation}
giving the PDF of equation (\ref{eq:puncsig}).

The strategy for the Gaussian model is identical, although the integrand in (\ref{eq:uncorpdf1}) is multiplied by the Gaussian model PDF (\ref{eq:p3}), with $\phi$ substituted by $\tau$.
The final Gaussian $\tau$ PDF therefore can be written in terms of a double integral, 
\begin{eqnarray}\lefteqn{
   P_{\mathrm{Gauss}}(\tau)  =  \frac{1500\sqrt{3} \eta}{\pi} \int_{0}^{1} d c_1 \int_{-1}^{1} d s_2 \times   }\label{eq:pgauss} \\
   &&\hspace{-1.0cm} \frac{a^4 (a^2 - \eta^2 s_1^2 (c_1 c_2 \eta + a s_2)^2) (a^2 - \eta^2 s_1^2 c_2^2)}
   {c_2 \left(\eta^2 s_1^2 (3 c_2^2 + (c_1 c_2 \eta + a s_2)(c_1 c_2 \eta + 3 a s_2)) + a^2 (5 + 3 c_1^2 \eta^2)\right)^{7/2}},
   \nonumber
\end{eqnarray}
where
\begin{equation}
   \eta = \sqrt{1-\tau^2},\;
   s_1 = \sqrt{1-c_1^2},\;
   c_2 = \sqrt{1-s_2^2},\;
   a = \sqrt{1- \eta^2 c_1^2}.
   \label{eq:settings}
\end{equation}
As in (\ref{eq:planarcond}) and Figure \ref{fig:planarprob}, negative values of $s_2$ appear as they replace values of $\phi$ between $\pi/2$ and $3\pi/2.$




\bsp

\label{lastpage}

\end{document}